\documentclass[12pt]{article}

\usepackage[totalheight = 23cm, totalwidth = 17cm]{geometry}
\usepackage{amssymb,amsmath,amsfonts,amsbsy,graphicx,bm}

\usepackage{color,cancel,ulem}

\newcommand{\dis}[1]{\begin{equation}\begin{split}#1\end{split}\end{equation}}

\begin{document}

\begin{titlepage}

\begin{center}

{\LARGE \bf 
Asymptotic behavior of saxion--axion system in stringy quintessence model
}

\vskip 1.0cm

{\large
Min-Seok Seo$^{a}$ 
}

\vskip 0.5cm

{\it
$^{a}$Department of Physics Education, Korea National University of Education,
\\ 
Cheongju 28173, Republic of Korea
}

\vskip 1.2cm

\end{center}

\begin{abstract}

 The late time behavior of the slow-roll parameter in the stringy quintessence model is studied when axion as well as saxion are allowed to move. Even though the potential is independent of the axion at tree level, the axion can move through its coupling to the saxion and the background geometry. Then the contributions of the axion kinetic energy to the slow-roll parameter and the vacuum energy density are not negligible when the slow-roll approximation does not hold. As the dimension of the field space is doubled, the fixed point at which the time variation of the slow-roll parameter vanishes is not always stable. It is found that the fixed point in the saxion--axion system is at most partially stable, in particular when the volume modulus and the axio-dilaton, the essential ingredients of the string compactification, are taken into account. It seems that as   more   saxion--axion pairs are considered, achieving the stability of the fixed point becomes difficult.

\end{abstract}

\end{titlepage}

\newpage

\section{Introduction}

Recently, there has been increasing interest in the question whether  the accelerating universe as we observe it today can be realized in the asymptotic regions of the moduli space, when the effective field theory characterized by the moduli space is   UV completed in string theory.
Whereas the metastable de Sitter (dS) vacuum is well compatible with the cosmological observations \cite{Planck:2018vyg},   the string models for it refs. \cite{Kachru:2003aw, Balasubramanian:2005zx} suffer from the parametric control issue.
That is, the parametric control in the asymptotic region is achieved when the potential is dominated by the small number of runaway terms, which however are required to be significantly corrected to realize the metastable dS vacuum \cite{Dine:1985he}.
 This can be well illustrated by  the difficulty in the stabilization of the K\"ahler moduli in  Type IIB string compactification on which the models \cite{Kachru:2003aw, Balasubramanian:2005zx} are based :
 unlike the complex structure moduli and the axio-dilaton, the K\"ahler moduli are not stabilized at tree level even if the fluxes are turned on due to the scale invariance of the internal manifold \cite{Giddings:2001yu}.
 Then to stabilize the K\"ahler moduli, corrections from the string length or  the non-perturbative effects need to be sizeable.
 Moreover, an appropriate size of the supersymmetry breaking uplift term is required in addition for the vacuum to have the positive energy density  and be sufficiently stable  \cite{Kachru:2002gs}.

As an alternative, an almost constant vacuum energy density driving the cosmic acceleration also can be explained by the quintessence model, in which the moduli slowly roll down the runaway potential with the small decay rate.
If it can be well realized in the framework of string theory,  we may account for the  accelerating universe compatible with the observations without loss of the parametric control (see refs.  \cite{Hellerman:2001yi, Fischler:2001yj, Kaloper:2008qs, Cicoli:2012tz} for earlier discussions and refs.  \cite{Agrawal:2018own, Cicoli:2018kdo, DavidMarsh:2018etu, Hebecker:2019csg, Olguin-Trejo:2018zun, Cicoli:2020cfj, ValeixoBento:2020ujr, Cicoli:2020noz, Cicoli:2021fsd, Cicoli:2021skd, Brinkmann:2022oxy, Conlon:2022pnx, Rudelius:2022gbz, Calderon-Infante:2022nxb, Apers:2022cyl, Shiu:2023nph, Shiu:2023fhb, Cremonini:2023suw, Hebecker:2023qke,  VanRiet:2023cca, Revello:2023hro, Apers:2024ffe, Seo:2024fki} for recent discussions concerning the parametric control).
However, it turns out that  when the unnatural corrections are suppressed such that  the K\"ahler moduli roll down the runaway potential, the decay rate of the potential is too large, or equivalently, the potential is too steep, to explain the almost constant vacuum energy density  \cite{Olguin-Trejo:2018zun, Cicoli:2021fsd}.
Such obstructions may indicate that our universe is a result of a nontrivial fine-tuning between the tree level runaway potential and various types of (perturbative as well as non-perturbative) corrections originating from either  quantum or stringy effects (see,   e.g., ref. \cite{ValeixoBento:2020ujr} for difficulties in realizing quintessence model even in the presence of corrections).

 The deviation of the spacetime geometry from dS space, the spacetime with the constant vacuum energy density, is directly measured by the Hubble slow-roll parameter 
 \dis{\epsilon=-\frac{\dot H}{H^2},}
 as the vacuum energy density is given by $3 M_{\rm Pl}^2 H^2$, where $H$ is the Hubble parameter (and also  the inverse of the horizon radius) and $M_{\rm Pl}$ is the reduced Planck mass.
 When the moduli slowly roll down the potential, $\epsilon$ can be approximated by the potential slow-roll parameter,  the slope of the potential in units of the Hubble parameter.
 For the runaway potential, the decay rate   is nothing more than the potential slow-roll parameter, hence it more or less well describes the deviation of the geometry from dS space in the slow-roll approximation. 
 This is, however,  no longer the case when the decay rate is of order one, then we need  to consider $\epsilon$ rather than the decay rate to appropriately describe the geometry \cite{Shiu:2023nph, Shiu:2023fhb, Seo:2024fki} (see also refs. \cite{Seo:2018abc, Mizuno:2019pcm} which point out this in terms of the inflationary model).
 Moreover,  the evolution of the moduli in this case cannot be well described by  the expansion with respect to $\epsilon$.
 We may compare this situation with the evolution of sizeable couplings under the renormalization group in  quantum field theory.
 When the (dimensionless) couplings are not much smaller than $1$, they cannot be trustable expansion parameters to describe the theory.
 Instead,  we may observe the behaviors of the couplings around the fixed point where the conformal invariance emerges.
 Motivated by this, we expect that the `fixed point values' of the moduli can be found and the behaviors of the moduli around the fixed points are determined by the stability of them.
 Recent analyses  show these aspects in detail, and there was an attempt to understand them in a simple manner using the language of the renormalization group \cite{Seo:2024fki}.  
The results of ref. \cite{Seo:2024fki} can be summarized as follows.
 If only the single modulus rolls down the potential and the decay rate is  smaller than some critical value (more concretely, the product of parameters  $\alpha$ and $\beta$   defined in \eqref{eq:action1} is smaller than $\sqrt6$), the stable fixed point value of $\epsilon$ coincides with  the potential slow-roll parameter, i.e., the decay rate of the potential, even if the slow-roll approximation does not hold.
 In contrast, if the decay rate of the  potential is larger than the critical value, $\epsilon$ converges  to $3$, the largest value of $\epsilon$ allowed by the positivity of the potential, regardless of the value of the decay rate.
 In the multifield case, the fixed point value of $\epsilon$ is not simply given by the sum of the fixed point values in the single field case, and also restricted to be equal to or smaller than $3$.
It is also notable that when the volume modulus (a  K\"ahler modulus determining the volume of the internal manifold) and the  dilaton roll down the potential simultaneously, the fixed point value of $\epsilon$ is larger than $1$.
Thus, dS space is rapidly destabilized and the quintessence model becomes incompatible with the observations.
 
 Meanwhile, the modulus in string theory typically corresponds to the scalar component of the  supermultiplet,  which also contains  the pseudoscalar component, an axion (see, e.g., Table 2 in ref.  \cite{Brinkmann:2022oxy} for the concrete string model realizations).
 We will call the modulus in this case the ``saxion."
 Even though the potential at tree level is independent of the axion by the shift symmetry, the axion can move as it couples to the saxion and the background geometry.
 When the decay rate of the potential is large enough that the slow-roll approximation does not hold, the contribution of the kinetic energy of the axion   to the vacuum energy density is not negligible.
 Then we expect the asymptotic  bound on $\epsilon$ in this case is different from that when the axion is assumed to be at rest, which will be explored in this article (see also refs. \cite{Cicoli:2021fsd, Revello:2023hro} for recent relevant discussions).
     For this purpose, we follow the analysis of ref. \cite{Seo:2024fki} (see also  refs. \cite{Christodoulidis:2019jsx, Cicoli:2020cfj, Cicoli:2020noz, Christodoulidis:2021vye} for similar analysis).
 In particular,  in order to test the stability, we investigate the time evolution of the small field fluctuations around the fixed point in detail.
     This has been used to analyze the stability of the fixed point under the renormalization group evolution in quantum field theory, and by applying this to the analysis of the evolution of (s)axion,  we hope to obtain the intuition on the effects of the axion without relying on the numerical analysis.  
Moreover, we use the fact that when the saxion couples to the potential  universally (indeed, the volume modulus and the dilaton correspond to this case),  dynamics is completely determined by  the time variations of the axion and the saxion.
  Then we can find these values at which $\epsilon$ becomes constant  and the equations of motion are well satisfied.
  As we will see, however, the fixed point is at most partially stable, i.e., stable with respect to the time evolution from some specific range of directions only.
  Moreover, as we take more saxion--axion pairs into account, the stability of the fixed point seems to be more difficult to achieve.


  The organization of this article is  as follows.
  In Sec. \ref{Sec:Dynamics}  we summarize the generic features of dynamics of the saxion--axion system with the runaway potential.
  In Sec. \ref{Sec:Fixed}, we consider the single pair of saxion--axion to test the stability of the fixed point  for different values of the decay rate of the potential.
  After addressing the fixed point in the presence of a number of  saxion--axion pairs in Sec. \ref{Sec:multi}, we conclude.

 \section{Dynamics of the saxion--axion system }
 \label{Sec:Dynamics}
 
 Throughout this article, we are interested in the tree level dynamics of the complex scalar field, the scalar (pseudoscalar) part of which is called the saxion (axion) and denoted by $u$ ($v)$. 
 Moreover, the potential depends only on the saxion  and exhibits the runaway behavior.
 Ignoring the quantum corrections like the non-perturbative effects, the action is given by
 \dis{S=\int d^4 x a(t)^3\Big(\frac{M_{\rm Pl}^2}{2\alpha^2}\frac{{\dot u}(t)^2+{\dot v}(t)^2}{u(t)^2}-\frac{V_0}{u(t)^\beta}\Big),\label{eq:action1}}
 where $\alpha$ and $\beta$ are positive numbers and $a(t)$ is the scale factor.
 In terms of the canonically normalized saxion  $\varphi=\frac{M_{\rm Pl}}{\alpha}\log u$, the potential is written as $V_0 e^{-\alpha\beta\frac{\varphi}{M_{\rm Pl}}}$, thus the product $\alpha\beta$ is interpreted as the decay rate of the potential.
Taking the Einstein--Hilbert action into account in addition, we obtain following equations of motion : 
\dis{&3 M_{\rm Pl}^2 H^2 =\frac{M_{\rm Pl}^2}{2\alpha^2}\frac{{\dot u}^2+{\dot v}^2}{u^2} +\frac{V_0}{u^\beta},
\\
&3 M_{\rm Pl}^2 H^2+2 M_{\rm Pl}^2\dot{H}=-\frac{M_{\rm Pl}^2}{2\alpha^2}\frac{{\dot u}^2+{\dot v}^2}{u^2} +\frac{V_0}{u^\beta},
\\
&\ddot{u}+3H\dot{u}-\frac{{\dot u}^2}{u}+\frac{{\dot v}^2}{u}-\frac{\alpha^2\beta}{M_{\rm Pl}^2}\frac{V_0}{u^{\beta+1}}=0,
\\
&\ddot{v}+3H\dot{v}-2\frac{{\dot u}{\dot v}}{u}=0,}
which can be rewritten as 
\dis{&\epsilon=\frac{1}{2\alpha^2}\Big[\Big(\frac{\dot u}{Hu}\Big)^2+\Big(\frac{\dot v}{Hu}\Big)^2\Big],
\\
&\frac{V}{ H^2 M_{\rm Pl}^2}=\frac{V_0/u^\beta}{ H^2 M_{\rm Pl}^2}=3-\epsilon,
\\
&\frac{\ddot u}{H^2 u}=-3\frac{\dot u}{Hu}+\Big(\frac{\dot u}{Hu}\Big)^2-\Big(\frac{\dot v}{Hu}\Big)^2+\alpha^2\beta\frac{V_0/u^\beta}{H^2 M_{\rm Pl}^2},
\\
&\frac{\ddot v}{H^2 u}=-3\frac{\dot v}{Hu}+2\frac{\dot u}{Hu}\frac{\dot v}{Hu}.}
They show that all dynamics can be described in terms of two variables
\dis{x=\frac{1}{\sqrt2\alpha}\frac{\dot u}{Hu},\quad\quad
y=\frac{1}{\sqrt2\alpha}\frac{\dot v}{Hu},}
 since $\frac{\ddot u}{H^2 u}$ and $\frac{\ddot v}{H^2 u}$ as well as $\frac{V}{ H^2 M_{\rm Pl}^2}$  depend only on them :  
 \dis{&\frac{\ddot u}{H^2 u}=-3\sqrt2 \alpha x+2\alpha^2x^2-2\alpha^2y^2 +\alpha^2\beta(3-\epsilon),
 \\
 &\frac{\ddot v}{H^2 u}=-3\sqrt2 \alpha y +4\alpha^2 xy.}
 Moreover, the positivity of the potential requires that
\dis{\epsilon=x^2+y^2 \leq 3.}

Meanwhile,  the time variation of   $\epsilon$  is given by
\dis{\frac{d\epsilon}{Hdt}&=\frac{1}{\alpha^2}\Big[\frac{\dot u}{Hu}\Big(\frac{\ddot u}{H^2 u}-\Big(\frac{\dot u}{Hu}\Big)^2+\epsilon \frac{\dot u}{Hu}\Big)+ 
\frac{\dot v}{Hu}\Big(\frac{\ddot v}{H^2 u}-\frac{\dot u}{Hu}\frac{\dot v}{Hu}+\epsilon \frac{\dot v}{Hu}\Big)\Big]
\\
&=2(\epsilon-3)\Big(\epsilon-\frac{\alpha\beta}{\sqrt2}x\Big),}
which shows that at the fixed point ($\frac{d\epsilon}{Hdt}=0$), at least one of two conditions 
\dis{x^2+y^2-\frac{\alpha\beta}{\sqrt2}x=0, \quad\quad x^2+y^2=3,\label{eq:fixedcond}}
is satisfied.
Indeed, since $\epsilon-3 \leq 0$, $\frac{d\epsilon}{Hdt}$ is positive when
\dis{x^2+y^2-\frac{\alpha\beta}{\sqrt2}x<0,\quad\quad x^2+y^2<3}
are satisfied simultaneously, and the   fixed points belong to the boundary of this region.
However, not all values of $(x, y)$ on the boundary can  be the   fixed point : we need to check the consistency with the equations of motion.
To be more concrete, we note that a pair of values $(x, y)$ in general varies in time as
\dis{\frac{dx}{Hdt}&=\frac{1}{\sqrt2\alpha}\frac{d}{Hdt}\Big(\frac{\dot u}{Hu}\Big)=\frac{1}{\sqrt2\alpha}\Big[\frac{\ddot u}{H^2u}-\Big(\frac{\dot u}{Hu}\Big)^2+\epsilon \frac{\dot u}{Hu}\Big]
\\
&=(\epsilon-3)\Big(x-\frac{\alpha\beta}{\sqrt2}\Big)-\sqrt2 \alpha y^2,
\\
\frac{d y}{Hdt}&=\frac{1}{\sqrt2\alpha}\frac{d}{Hdt}\Big(\frac{\dot v}{Hu}\Big)=\frac{1}{\sqrt2\alpha}\Big[\frac{\ddot v}{H^2u}- \frac{\dot u}{Hu}\frac{\dot v}{Hu}+\epsilon \frac{\dot v}{Hu}\Big]
\\
&=(\epsilon-3)y+\sqrt2 \alpha x y.
}
Then   three cases obeying the fixed point condition $\frac{d\epsilon}{Hdt}=0$ are more restricted by time variations  of $x$ and $y$ as follows : 

\begin{itemize}
\item {\bf Case 1} : $\epsilon=\frac{\alpha\beta}{\sqrt2}x$ but $\epsilon < 3$.
 Since $\epsilon$ is a constant in time at the fixed point, the relation $\epsilon=\frac{\alpha\beta}{\sqrt2}x$ imposes that $x$ (hence $y$) is also a constant in time, i.e., $\frac{dx}{Hdt}=\frac{dy}{Hdt}=0$.
Meanwhile, the curves $x^2+y^2-\frac{\alpha\beta}{\sqrt2}x=0$ and  $x^2+y^2=\epsilon$ intersect   at 
\dis{x=\frac{\sqrt2 \epsilon}{\alpha\beta},\quad \quad y=\sqrt\epsilon \sqrt{1-\frac{2\epsilon}{(\alpha\beta)^2}},\label{eq:case1int}}
 for nonzero  $\alpha$ and $\beta$,  then the condition $\frac{dx}{Hdt}=\frac{dy}{Hdt}=0$ is satisfied provided
 \dis{\epsilon=\frac{3\beta}{2+\beta},\quad\quad{\rm or}\quad\quad \epsilon=\frac{(\alpha\beta)^2}{2}.}
 We note that $\epsilon=\frac{3\beta}{2+\beta}$ is compatible with the positivity of the potential since it is always smaller than $3$.
 Moreover, the slow-roll condition $\epsilon<1$ is satisfied when $\beta<1$.
 Meanwhile, $\epsilon=\frac{(\alpha\beta)^2}{2}$ is consistent with the positivity of the potential only if $\alpha\beta \leq \sqrt6$.
 In this case, the value of $y$ is fixed to zero. 
 \item {\bf Case 2} : $\epsilon \ne \frac{\alpha\beta}{\sqrt2}x$ but $\epsilon = 3$.
  In this case,  time variations of $x$ and $y$ at the fixed point satisfy
 \dis{\frac{dx}{Hdt}=-\sqrt2 \alpha y^2,\quad\quad \frac{dy}{Hdt}=\sqrt2 \alpha x y.\label{eq:case2xyder}}
That is, while  $x$ ($y$) decreases (increases)  in time, the value of $\epsilon$ is fixed to $3$.
\item {\bf Case 3} : $\epsilon=\frac{\alpha\beta}{\sqrt2}x$ and $\epsilon = 3$.
 Whereas the values of   $x$ and $y$ at the fixed point are given by
\dis{x=\frac{3\sqrt2}{\alpha\beta},\quad\quad y=\sqrt{3-\frac{18}{(\alpha\beta)^2}},\label{eq:case3xy}}
their time variations 
\dis{\frac{dx}{Hdt}=-\sqrt2\alpha\Big(3-\frac{18}{(\alpha\beta)^2}\Big),\quad\quad \frac{dy}{Hdt}=\frac{6}{\beta}\sqrt{3-\frac{18}{(\alpha\beta)^2}}\label{eq:case3xyder}}
vanish only if $\alpha\beta=\sqrt6$, indicating that $y$ is fixed to zero.
This is nothing more than the continuation of $\epsilon=\frac{(\alpha\beta)^2}{2}$ in Case 1.
\end{itemize}

The stability of the fixed point is determined by not only the values of $\alpha$ and $\beta$, but also the behaviors of $x$ and $y$ under the perturbation, which will be discussed in the next section.
Before closing this section, we note that we restrict our attention to the positive $x=\frac{1}{\sqrt2 \alpha}\frac{\dot u}{Hu}$, in which the value of the saxion keeps increasing   in time, i.e., the saxion rolls down, rather than climbs up the potential.
Whereas the value of $y=\frac{1}{\sqrt2 \alpha}\frac{\dot v}{Hu}$ can be negative, since the action is invariant under $v\to -v$,  our discussion on the positive $y$ also applies to $-y$.
For this reason, we consider the positive values of $x$ and $y$.

 \section{Stability of the fixed point }
  \label{Sec:Fixed}
 
 In this section, we investigate the late time behavior of $\epsilon$ more carefully by considering the time evolutions of $x$ and $y$ for different values of     $\alpha$ and  $\beta$. 
 If we consider the dynamics of the saxion only,  the time variation of $\epsilon$ depends only on $x$, which is nothing more than $\sqrt{\epsilon}$.
 Then the boundary between the regions of the positive and the negative $\frac{d\epsilon}{Hdt}$ in the one-dimensional field space $x$ is easily interpreted as the stable fixed point : we cannot cross the fixed point through the time evolution when $\frac{d\epsilon}{Hdt}$ is a function of the single variable $\epsilon=x^2$. 
 However, by taking the axion into account in addition, the time variation of $\epsilon$ is considered in two-dimensional field space, $(x, y)$.
 In this case,  we have another direction which allows the point $(x, y)$ not to go back to the fixed point after it evolves across the fixed point to change the sign of $\frac{d\epsilon}{Hdt}$.   
 Thus  the fixed point may be stable with respect to the time evolution from  some specific range of directions only.

 \subsection{$\alpha\beta \leq \sqrt6$}
 
 \begin{figure}[!t]
 \begin{center}
 \includegraphics[width=0.45\textwidth]{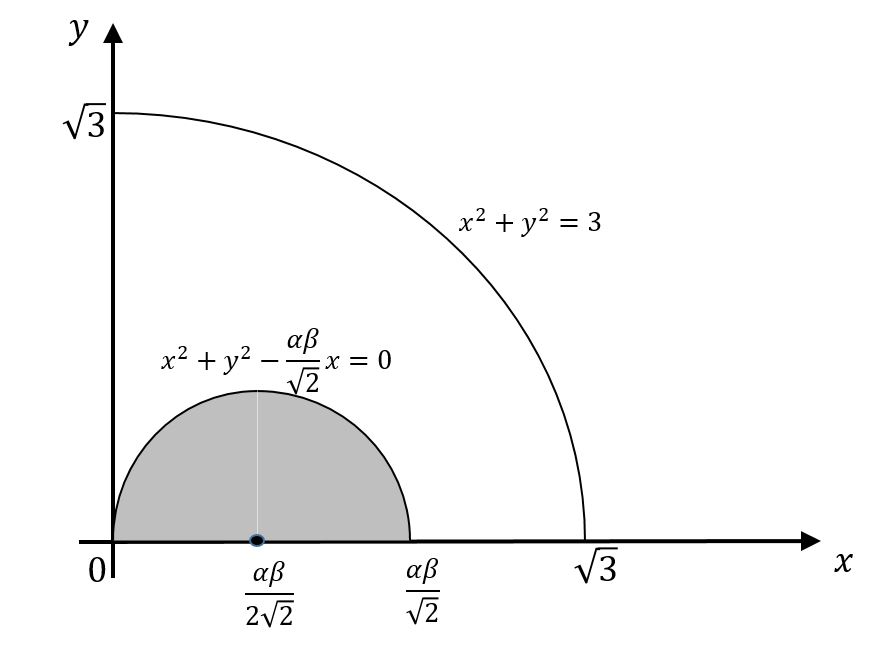}
 \includegraphics[width=0.35\textwidth]{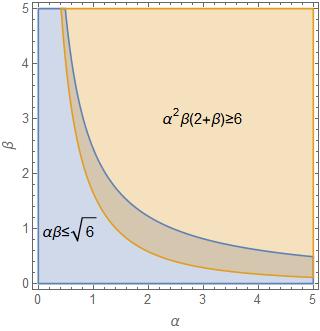}
 \end{center}
\caption{ Left : two curves $x^2+y^2=3$ and $x^2+y^2-\frac{\alpha\beta}{\sqrt2}x=0$ when $\alpha\beta<\sqrt6$.
The value of $\frac{d\epsilon}{Hdt}$ is positive in the region $x^2+y^2-\frac{\alpha\beta}{\sqrt2}x<0$ which is colored in gray, while negative in the region between two curves, the overlapping region between $x^2+y^2<3$ and $x^2+y^2-\frac{\alpha\beta}{\sqrt2}x>0$.
Right : overlapping region between $\alpha\beta\leq \sqrt6$ and $\alpha^2\beta(2+\beta)\geq 6$, which allows the fixed point value $\epsilon=\frac{3\beta}{2+\beta}$.
  }
\label{Fig:abss6}
\end{figure}
  
 When $\alpha\beta<\sqrt6$, $\frac{d\epsilon}{Hdt}$ is positive in the region 
\dis{x^2+y^2-\frac{\alpha\beta}{\sqrt2}x<0}
(colored in gray in the left panel of Fig. \ref{Fig:abss6}), the whole of which is inside the region $x^2+y^2 < 3$ (or equivalently, $V>0$) as depicted in the left panel of Fig. \ref{Fig:abss6}.
Since $\frac{d\epsilon}{Hdt}$ is negative in the overlapping region between   $x^2+y^2<3$ and $x^2+y^2-\frac{\alpha\beta}{\sqrt2}x>0$, the point $(x, y)$ on the curve $x^2+y^2=3$, at which $\frac{d\epsilon}{Hdt}=0$ and $\epsilon=3$, corresponds to the unstable fixed point.
Meanwhile, as can be found in the discussion on  Case 1 in Sec. \ref{Sec:Dynamics}, the values of $\epsilon$ as well as  $(x, y)$ at the  fixed points satisfying $x^2+y^2-\frac{\alpha\beta}{\sqrt2}x=0$ are given as follows:
\dis{\epsilon=\frac{3\beta}{2+\beta}\quad &{\rm at}\quad x=\frac{3\sqrt2}{\alpha(2+\beta)},\quad y=\frac{\sqrt{3(\alpha^2\beta(2+\beta)-6)}}{\alpha(2+\beta)},
\\
\epsilon=\frac{(\alpha\beta)^2}{2}\quad &{\rm at}\quad x=\frac{\alpha\beta}{\sqrt2},\quad y=0.\label{eq:case1<6}}
It is  obvious from \eqref{eq:case1int} that the values of $x$ and $y$ at the fixed point are physically sensible provided $\epsilon \leq \frac{(\alpha\beta)^2}{2}$, which can be found from the fact that the value of $x$ on the curve $x^2+y^2-\frac{\alpha\beta}{\sqrt2}x=0$ is smaller than $\frac{\alpha\beta}{\sqrt2}$ where $\epsilon = \frac{(\alpha\beta)^2}{2}$ (see Fig. \ref{Fig:abss6}). 
In particular, for $(x, y)$ corresponding to the fixed point value $\epsilon=\frac{3\beta}{2+\beta}$ to exist, $\alpha$ and $\beta$ must satisfy $\alpha^2\beta(2+\beta) \geq 6$.
The values of $\alpha$ and $\beta$   belonging to the overlapping region between $\alpha\beta <\sqrt6$ and $\alpha^2\beta(2+\beta) \geq 6$ are shown in   the right panel of Fig. \ref{Fig:abss6}.
For the axio-dilaton   in Type IIB string compactification, $(\alpha, \beta)=(\sqrt2, 1)$, hence $\alpha\beta =\sqrt2 <\sqrt 6$ and $\alpha^2\beta(2+\beta)=6$.
In this case, two fixed points coincide, i.e.,
\dis{\epsilon=\frac{3\beta}{2+\beta}=\frac{(\alpha\beta)^2}{2}=1,}
at $(x, y)=(1,0)$.

   Meanwhile, when $\alpha\beta=\sqrt6$, two curves $x^2+y^2 = 3$ and $x^2+y^2-\frac{\alpha\beta}{\sqrt2}x= 0$ intersect at $(x, y)=(\sqrt3,0)$ giving $\epsilon=3$, which corresponds to  Case 3.
 Moreover, the point $(x, y)$ satisfying $\epsilon=\frac{\alpha\beta}{\sqrt2}x$ except for $(\sqrt3, 0)$ corresponds to  Case 1, where the  fixed point value is given by the first line in \eqref{eq:case1<6}.
 An example for this case is the volume modulus in Type IIB string compactification :  since  $(\alpha, \beta)=(\sqrt{\frac23}, 3)$, we can find two fixed points $\epsilon=3$ at $(x, y)=(\sqrt3, 0)$ and $\epsilon=\frac95$ at $(x, y)=(\frac{3\sqrt3}{5}, \frac{3\sqrt2}{5})$.

  Now, we investigate the stability of the fixed points we found above.
   For the fixed point to be stable, the point $(x, y)$ with  $\epsilon$ smaller (larger) than the fixed point value (which will be denoted by $\epsilon_0$) in the region $\frac{d\epsilon}{Hdt}>0$ ($\frac{d\epsilon}{Hdt}<0$) must evolve into the fixed point.
   To see this, we consider the values of $x$ and $y$ slightly deviate from the fixed point values and see their time variations $\frac{dx}{Hdt}$ and $\frac{dy}{Hdt}$. 
 
\subsubsection{Case 1 with $\epsilon_0=\frac{3\beta}{2+\beta}$}
  
  For the fixed point corresponding to $\epsilon_0=\frac{3\beta}{2+\beta}$ (the first case in \eqref{eq:case1<6}), the small deviations around the fixed point 
  \dis{&x=\frac{3\sqrt2}{\alpha(2+\beta)}+\delta x\equiv x_0+\delta x
  \\
&y=\frac{\sqrt{3(\alpha^2\beta(2+\beta)-6)}}{\alpha(2+\beta)}+\delta y \equiv y_0+\delta y}
  give 
  \dis{&\frac{dx}{Hdt}=-\frac{6(\alpha^2(\beta^2+3\beta+2)-6)}{\alpha^2(2+\beta)^2}\delta x -\frac{\sqrt{6(\alpha^2\beta(2+\beta)-6)}(\alpha^2(\beta+2)^2-6)}{\alpha^2(2+\beta)^2}\delta y,
  \\
  &\frac{dy}{Hdt}=\frac{\sqrt{6(\alpha^2\beta(2+\beta)-6)}(\alpha^2(\beta+2)+6)}{\alpha^2(2+\beta)^2}\delta x + \frac{6(\alpha^2\beta(2+\beta)-6)}{\alpha^2(2+\beta)^2}\delta y.}
Since $\alpha^2\beta(2+\beta)> 6$, the coefficients of $\delta x$ and $\delta y$ in $\frac{dx}{Hdt}$ are negative while those in $\frac{dy}{Hdt}$ are positive.
Then near the fixed point, the condition $\frac{dx}{Hdt}>0$ ($\frac{dx}{Hdt}<0$) can be written as $\delta y<-k_x \delta x$ ($\delta y > -k_x \delta x$) where
\dis{k_x=\frac{\sqrt6 (\alpha^2(\beta^2+3\beta+2)-6) }{\sqrt{ (\alpha^2\beta(2+\beta)-6)}(\alpha^2(\beta+2)^2-6)},}
 while the condition $\frac{dy}{Hdt}>0$ ($\frac{dy}{Hdt}<0$) can be written as $\delta y>-k_y \delta x$ ($\delta y < -k_y \delta x$) where
\dis{k_y=\frac{\alpha^2(\beta+2)+6}{\sqrt{6(\alpha^2\beta(2+\beta)-6)}}.}
Here $k_x$ and $k_y$ are both positive but do not coincide.
In fact, for any positive $\alpha$ and $\beta$ satisfying $\alpha^2\beta(2+\beta)-6>0$, $k_x$ is always smaller than $k_y$.
We will compare them with the tangent line of $x^2+y^2=\epsilon_0=$(constant) and that of $x^2+y^2-\frac{\alpha\beta}{\sqrt2}x=0$ at the fixed point, which are given by $\delta y=-c_0 \delta x$, where 
\dis{c_0= \frac{\sqrt6}{\sqrt{ (\alpha^2\beta(2+\beta)-6)}} }
and  $\delta y=-c_1 \delta x$,  where
\dis{c_1=-\frac{\alpha^2\beta(2+\beta)-12}{2\sqrt6 \sqrt{ (\alpha^2\beta(2+\beta)-6)} }.}
respectively.
We note that whereas $c_0$ is positive, $c_1$ is positive (negative) if  $x_0$ is larger (smaller) than $\frac{\alpha\beta}{2\sqrt2}$, the $x$ value of the center of the circle $x^2+y^2-\frac{\alpha\beta}{\sqrt2}x=0$, which is obvious from Fig. \ref{Fig:abss6}.
Moreover, when $c_1$ is positive, one immediately finds that $c_0>c_1$.
While $|c_1|$ may be larger than $c_0$ for a negative $c_1$, it is not relevant to our discussion.

 \begin{figure}[!t]
 \begin{center}
  \includegraphics[width=0.45\textwidth]{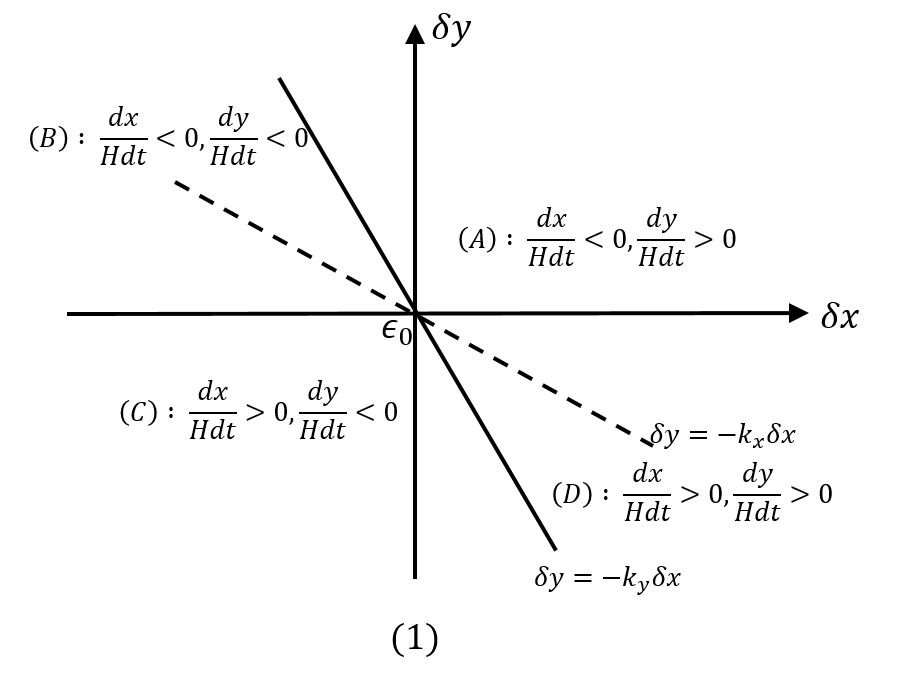}
  \\
 \includegraphics[width=0.45\textwidth]{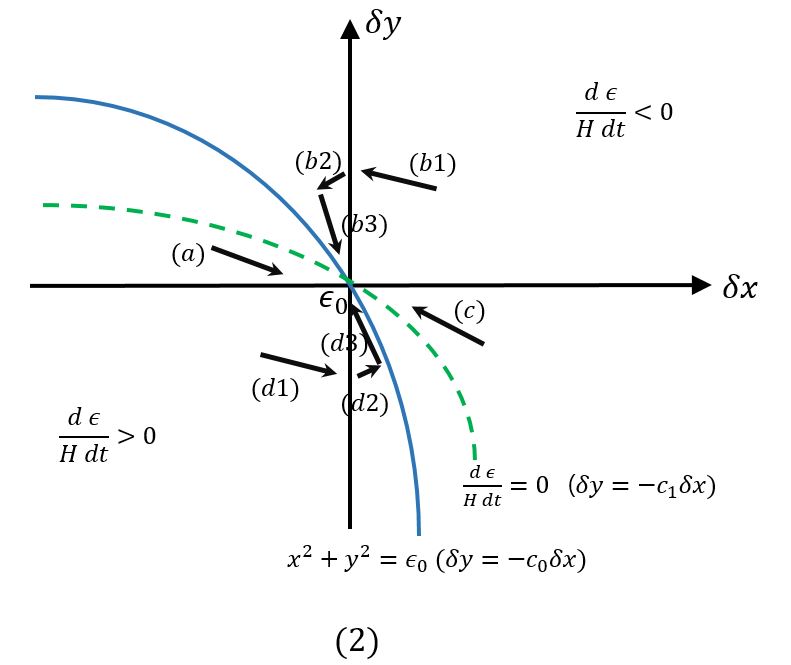}
  \includegraphics[width=0.45\textwidth]{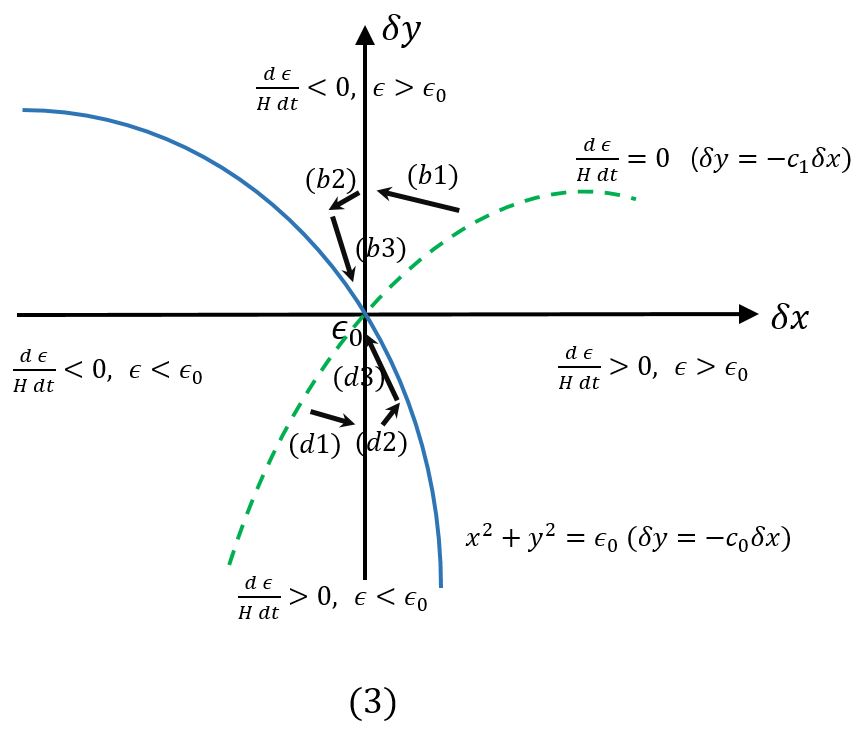}
 \end{center}
\caption{  (1)    The regions near the fixed point $\epsilon_0$ distinguished by the signs of $\frac{dx}{Hdt}$ and $\frac{dy}{Hdt}$.
Here $\frac{dx}{Hdt}=0$ on the black dashed line $\delta y=-k_x \delta x$ and $\frac{dy}{Hdt}=0$ on the black  line $\delta y=-k_y \delta x$. We note that $k_x<k_y$.
(2)  Possible time evolutions of $(x, y)$ when $\alpha^2\beta(2+\beta)<12$. The arrows indicate the time evolutions of $(x, y)$ near the fixed point when the fixed point is stable. 
The blue line and the green dashed line indicate the curves $x^2+y^2=\epsilon_0$ (approximated by $\delta y=-c_0\delta x$) and $\frac{d\epsilon}{Hdt}=0$ (or equivalently, $x^2+y^2-\frac{\alpha\beta}{\sqrt2}x=0$ : approximated by $\delta y=-c_1\delta x$), respectively.
(3) Possible time evolutions of $(x, y)$ when $\alpha^2\beta(2+\beta)\geq 12$.
  }
\label{Fig:abss6sta}
\end{figure}

 From observations so far, we investigate the stability of the fixed point. 
 Since $k_x<k_y$, the region  around the fixed point can be divided according to the signs of $\frac{dx}{Hdt}$ and $\frac{dy}{Hdt}$ as shown in   Fig. \ref{Fig:abss6sta} (1). 
 We first consider the case $\alpha^2\beta (2+\beta)<12$, that is,  $x_0$ is larger than $\frac{\alpha\beta}{2\sqrt2}$. 
 When  the point $(x, y)=(x_0+\delta x, y_0+\delta y)$ satisfies both $\epsilon<\epsilon_0$ and  $\frac{d\epsilon}{Hdt}>0$,  it belongs to the overlapping region between $\delta y < -c_0 \delta x$ and $\delta y <-c_1 \delta x$  for sufficiently small values of $\delta x$ and $\delta y$ (see Fig. \ref{Fig:abss6sta} (2)). 
If $\delta x<0$ and  $\delta y>0$ for $(x, y)$ in this case, the fixed point is stable with respect to the time evolution of $(x, y)$ provided  $\frac{dx}{Hdt}>0$ and $\frac{dx}{Hdt}<0$ are satisfied,  i.e.,   $(x,  y)$ belongs  to the region (C) in  Fig. \ref{Fig:abss6sta} (1) and evolves following the arrow (a) in   Fig. \ref{Fig:abss6sta} (2).
 Indeed, since   $c_1<k_x$, this condition  is trivially fulfilled for any $\alpha$ and $\beta$ satisfying $\alpha^2\beta(2+\beta)-6>0$ (otherwise, $y_0$ is not well defined : see \eqref{eq:case1<6}).
 Moreover, the relation $c_1<k_x$ also allows the stability of the fixed point with respect to the   evolution  of $(x, y)$ with $\delta x > 0$ and $\delta y<0$ when it  is in the  overlapping region between $\delta y>-c_0 \delta x$ and $\delta y>-c_1 \delta x$.
 In this case, $(x, y)$ satisfies $\epsilon>\epsilon_0$ and $\frac{d\epsilon}{Hdt}<0$ and its time evolution follows  the arrow (c) in   Fig. \ref{Fig:abss6sta} (2). 
  Meanwhile, if $\delta x<0$ and $\delta y<0$, since $\frac{dx}{Hdt}>0$ and $\frac{dy}{Hdt}<0$, $(x, y)$ moves away from the fixed point, following the arrow (d1). 
 However, if $(x, y)$ in this case can evolve into the region $(\frac{dx}{Hdt}>0, \frac{dy}{Hdt}>0)$ (the region (D)) and then   into the region $(\frac{dx}{Hdt}<0, \frac{dy}{Hdt}>0)$ (the region (A)) following the arrows (d1)$\to$(d2)$\to$(d3), it can reach the fixed point.
   Since all these processes take place satisfying $\frac{d\epsilon}{Hdt}>0$ (that is, $\epsilon$ increases in time), $\epsilon$ must be always smaller than $\epsilon_0$, which imposes the condition $c_0<k_x$.
    But this is not satisfied for any $\alpha$ and $\beta$ satisfying $\alpha^2\beta(2+\beta)-6>0$ so the fixed point is not stable under the evolution of $(x, y)$ in the region $\delta y<0$ and $\epsilon<\epsilon_0$.
   The same argument is used to show that the fixed point is not stable for $(x, y)$ satisfying $\delta y>0$ and $\epsilon>\epsilon_0$ : the arrow (b3) is not allowed.
   In summary, when $\alpha^2\beta(2+\beta)<12$, the fixed point is  stable with respect to the time evolution of $(x, y)$ only if
   \dis{&\delta x<0, ~\delta y>0 \quad {\rm for}~\epsilon<\epsilon_0~{\rm and}~ \frac{d\epsilon}{Hdt}>0,
   \\
   &\delta x>0, ~\delta y<0 \quad {\rm for}~\epsilon>\epsilon_0~{\rm and}~ \frac{d\epsilon}{Hdt}<0.} 
 Indeed, the volume modulus in Type IIB string compactification corresponds to this case : from $(\alpha, \beta)=(\sqrt{\frac23}, 3)$ we find that  $\alpha^2\beta(2+\beta)=10$ is larger than $6$ but smaller than $12$ and the slopes appearing in our discussion are given by  $c_0=\sqrt{\frac32}$, $c_1=\frac{1}{2\sqrt6}$, $k_x=\frac{11}{16}\sqrt{\frac32}$, and  $k_y=\frac73\sqrt{\frac23}$,  respectively. 
 Therefore, the fixed point $\epsilon=\frac95$ at $(x, y)=(\frac{3\sqrt3}{5}, \frac{3\sqrt2}{5})$ is partially stable.

   The stability of the fixed point in the case of $\alpha^2\beta(2+\beta)\geq 12$, i.e., $x_0 \leq \frac{\alpha\beta}{2\sqrt2}$, can be tested in the same manner.   
When the point $(x, y)$ is located in the region $\frac{d\epsilon}{Hdt}<0$ and $\epsilon>\epsilon_0 $, it can reach the fixed point provided it evolves into the region $(\frac{dx}{Hdt}>0, \frac{dy}{Hdt}<0)$ (the region (C)), where the time evolution of $(x, y)$ follows an arrow (b3) in Fig. \ref{Fig:abss6sta} (3).
 This requires $c_0< k_x$ which however is not satisfied for any $\alpha$ and $\beta$ satisfying $\alpha^2\beta(2+\beta)-6>0$.
 Consideration of the evolution of  $(x, y)$ in the region $\frac{d\epsilon}{Hdt}>0$ and $\epsilon<\epsilon_0$  also gives the same conclusion.
 Thus, the fixed point in the region $\alpha^2\beta(2+\beta)\geq 12$ is not stable.

\subsubsection{Case 1 with $\epsilon_0=\frac{(\alpha\beta)^2}{2}$, Case 3}

We now consider the fixed point corresponding to  $\epsilon_0=\frac{(\alpha\beta)^2}{2}$, the perturbation around which is given by 
 \dis{x=\frac{\alpha\beta }{\sqrt2} +\delta x,\quad y=\delta y.}
In this case, in the region $\delta x<0$ ($\delta x>0$), the value of $\epsilon$ is smaller (larger) than the fixed point value and  $\frac{d\epsilon}{Hdt}>0$ ($\frac{d\epsilon}{Hdt}<0$) is satisfied.
The time variations of $x$ and $y$ reads
  \dis{&\frac{dx}{Hdt}=-\Big(3-\frac{(\alpha\beta)^2}{2}\Big)\delta x,
  \\
  &\frac{dy}{Hdt}=\frac12 (\alpha^2\beta(\beta+2)-6)\delta y.}
Since the coefficient of $\delta x$ in $\frac{dx}{Hdt}$ is negative for $\alpha\beta<\sqrt6$, the fixed point is stable   provided $\alpha^2\beta(\beta+2)<6$ such that the coefficient of $\delta y$ in $\frac{dy}{Hdt}$ is also negative.
 Meanwhile, when $\alpha\beta=\sqrt6$, which corresponds to Case 3, the coefficient of $\delta x$ in $\frac{dx}{Hdt}$ vanishes whereas the coefficient of $\delta y$ in $\frac{dy}{Hdt}$ is given by the positive number $\sqrt6\alpha$. 
 This shows that the value of $y$ tends to move away from the fixed point value hence the fixed point in this case is not stable as well.
  Applying this to  Type IIB string compactification, one finds that the fixed point of the axio-dilaton, $(x, y)=(1,0)$ giving $\epsilon=1$  and that of the volume modulus,   $(x, y)=(\sqrt3,0)$ giving $\epsilon=3$  are unstable.

\subsection{$\alpha\beta>\sqrt6$}

 \begin{figure}[!t]
 \begin{center}
 \includegraphics[width=0.45\textwidth]{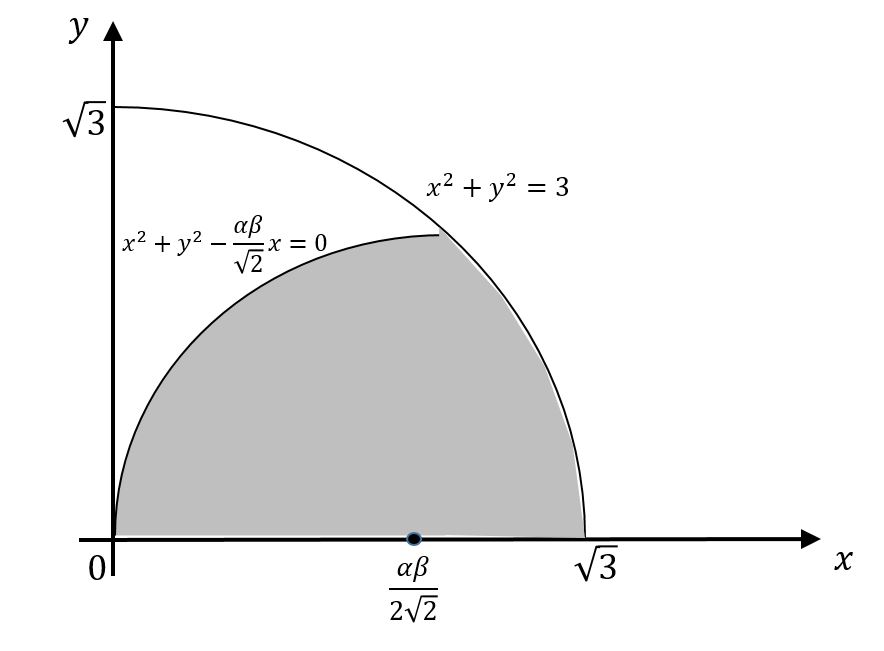}
 \end{center}
\caption{ Two curves $x^2+y^2=3$ and $x^2+y^2-\frac{\alpha\beta}{\sqrt2}x=0$ when $\alpha\beta>\sqrt6$.
The value of $\frac{d\epsilon}{Hdt}$ is positive in the overlapping region between $x^2+y^2<3$ and  $x^2+y^2-\frac{\alpha\beta}{\sqrt2}x<0$, which is colored in gray, while negative in its complement region, i.e., the overlapping region between   $x^2+y^2<3$ and  $x^2+y^2-\frac{\alpha\beta}{\sqrt2}x>0$.
  }
\label{Fig:absl6}
\end{figure}
 
 When $\alpha\beta>\sqrt6$, only a part of the region $x^2+y^2-\frac{\alpha\beta}{\sqrt2}x <0$ overlaps with $x^2+y^2 \leq 3$, and the values of $(x, y)$ belonging to this overlapping region (colored in gray in Fig. \ref{Fig:absl6}) satisfy $\frac{d\epsilon}{Hdt}>0$.
 Then we expect that the stable fixed point belongs to the boundary of the region $\frac{d\epsilon}{Hdt}>0$ consisting of a part of the curve $x^2+y^2-\frac{\alpha\beta}{\sqrt2}x =0$ and that of the curve $x^2+y^2 = 3$.
 The intersection point between two curves corresponds to Case 3, at which $\epsilon=3$ and a pair of values $(x, y)$ is given by \eqref{eq:case3xy}, but as we have seen, both $\frac{dx}{Hdt}$ and $\frac{dy}{Hdt}$ do not vanish there.
 Indeed, since $\frac{dx}{Hdt}<0$ and  $\frac{dy}{Hdt}>0$ at the intersection (see  \eqref{eq:case3xyder}), the point $(x, y)$ in this case   eventually evolves into the part of the curve $x^2+y^2=3$ which does not belong to the boundary of the region $\frac{d\epsilon}{Hdt}>0$.
 In fact, once $\epsilon$ becomes $3$, $(x, y)$ keeps rotating around the circle $x^2+y^2=3$, even allowing the negative values of $x$ or $y$ (that is, the saxion may climb up the potential).
 The part of the curve $x^2+y^2=3$ belonging to the boundary of the region $\frac{d\epsilon}{Hdt}>0$ corresponds to Case 2, in which the value   $\epsilon=3$ does not change in time despite   the time evolution of $(x, y)$.
As can be found in \eqref{eq:case2xyder}, $\frac{dx}{Hdt}<0$ and  $\frac{dy}{Hdt}>0$ in this case, which means that the point $(x, y)$ satisfying $\epsilon=3$ keeps rotating on  the curve $x^2+y^2=3$  as mentioned above. 
Meanwhile, the fixed point on a part of the curve $x^2+y^2-\frac{\alpha\beta}{\sqrt2}x =0$ belonging to the boundary of the region $\frac{d\epsilon}{Hdt}>0$ corresponds to   Case 1.
In this case, $\epsilon=\frac{3\beta}{2+\beta}$ and the values of $x$ and $y$ are given by \eqref{eq:case1int}.
While \eqref{eq:case1int} requires $(\alpha\beta)^2 \geq 2\epsilon$ (or equivalently, $\alpha^2\beta(2+\beta)> 6$), this condition is trivially fulfilled when $\alpha\beta>\sqrt6$ but $\epsilon <3$.
The argument on the stability under the perturbation is the same as that in the case of $\alpha\beta \leq \sqrt6$, so we conclude that the fixed point is partially stable provided   $\alpha^2\beta(2+\beta)< 12$.
Another fixed point corresponding to Case 1, $\epsilon=\frac{(\alpha\beta)^2}{2}$ is not allowed since it contradicts to the positivity of the potential ($\epsilon<3$) when $\alpha\beta>\sqrt6$.

\section{Fixed point in the multifield model}
\label{Sec:multi}

In the presence of multi-pair of saxions and axions  $(u_i, v_i)$, the action is given by
 \dis{S=\int d^4 x a(t)^3\Big(\sum_i \frac{M_{\rm Pl}^2}{2\alpha_i ^2}\frac{{\dot u_i}(t)^2+{\dot v_i}(t)^2}{u_i(t)^2}-V \Big),}
 where the tree level  potential $V$ does not depend on axions $v_i$, 
 from which we obtain the following equations of motion :
\dis{&3 M_{\rm Pl}^2 H^2 =\sum_i\frac{M_{\rm Pl}^2}{2\alpha_i^2}\frac{{\dot u_i}^2+{\dot v_i}^2}{u_i^2} +V,
\\
&3 M_{\rm Pl}^2 H^2+2 M_{\rm Pl}^2\dot{H}=-\sum_i \frac{M_{\rm Pl}^2}{2\alpha_i^2}\frac{{\dot u_i}^2+{\dot v_i}^2}{u_i^2} +V,
\\
&\ddot{u_i}+3H\dot{u_i}-\frac{{\dot u_i}^2}{u_i}+\frac{{\dot v_i}^2}{u_i}+\frac{\alpha_i^2}{M_{\rm Pl}^2}u_i^2\frac{\partial V}{\partial u_i }=0,
\\
&\ddot{v_i}+3H\dot{v_i}-2\frac{{\dot u_i}{\dot v_i}}{u_i}=0.}
We are interested in the volume modulus and the axio-dilaton in Type IIB string compactification which couple to the potential universally (that is, every   F-term potential term  depends on them in the same way), such that the potential satisfies $u_i \frac{\partial V}{\partial u_i}=-\beta_i V$.
Then in terms of
\dis{x_i=\frac{1}{\sqrt2 \alpha_i}\frac{{\dot u_i}}{Hu_i},\quad\quad y_i=\frac{1}{\sqrt2 \alpha_i}\frac{{\dot v_i}}{Hu_i},}
 the relations
\dis{&\epsilon=\sum_i(x_i^2+y_i^2),
\\
&\frac{V}{M_{\rm Pl}^2H^2}=3-\epsilon,
\\
&\frac{\ddot u}{H^2 u_i}=-3\sqrt2 \alpha_i x_i +2\alpha_i^2 x_i^2-2\alpha_i^2 y_i^2+\alpha_i^2\beta_i(3-\epsilon),
\\
&\frac{\ddot v}{H^2 u_i}=-3\sqrt2 \alpha_i y_i +4\alpha_i^2 x_i  y_i.}
are obeyed, then we obtain 
\dis{&\frac{dx_i}{Hdt}=\frac{1}{\sqrt2\alpha_i}\Big[\frac{\ddot u_i}{H^2u_i}-\Big(\frac{\dot u_i}{Hu_i}\Big)^2+\epsilon\frac{\dot u_i}{Hu_i}\Big]=(\epsilon-3)\Big(x_i-\frac{\alpha_i\beta_i}{\sqrt2}\Big)-\sqrt2 \alpha_i y_i^2,
\\
&\frac{dy_i}{Hdt}=\frac{1}{\sqrt2\alpha_i}\Big[\frac{\ddot v_i}{H^2u_i}- \frac{\dot u_i}{Hu_i}\frac{\dot v_i}{Hu_i}+\epsilon\frac{\dot v_i}{Hu_i}\Big]=(\epsilon-3)y_i+\sqrt2 \alpha_i x_iy_i.}

At the fixed point, the value of $\epsilon$ does not vary in time, i.e., 
\dis{\frac{d\epsilon}{Hdt}=2(\epsilon-3)\Big(\epsilon-\sum_i\frac{\alpha_i\beta_i}{\sqrt2}x_i\Big)=0.}
We first consider the case $\epsilon=\sum_i\frac{\alpha_i\beta_i}{\sqrt2}x_i$, which corresponds to the generalization of Cases 1 and 3 in Sec. \ref{Sec:Dynamics}.
At first glance, the condition that the combination $\sum_i\frac{\alpha_i\beta_i}{\sqrt2}x_i$  which is identified with $\epsilon$  is a constant in time  is less restrictive  than  the single field case ($\frac{dx_i}{Hdt}=\frac{dy_i}{Hdt}=0$), but it is not always the case.
To see this, we consider the simplest case in which two saxion-axion pairs   roll down the potential, i.e., the index $i$ runs from $1$ to $2$.
Since the combination $\sum_i\frac{\alpha_i\beta_i}{\sqrt2}x_i$ is a constant in time, we obtain another condition,  
\dis{\frac{d}{Hdt}\sum_i\frac{\alpha_i\beta_i}{\sqrt2}x_i = (\epsilon-3)\Big(\epsilon-\sum_i\frac{(\alpha_i\beta_i)^2}{2}\Big)-\sum_i\alpha_i^2\beta_i y_i^2 =0, \label{eq:multiCond3}}
 or equivalently, the combination $\sum_i\alpha_i^2\beta_i y_i^2$ is also a constant in time.
Then from 
\dis{\frac{d}{Hdt}\sum_i\alpha_i^2\beta_i y_i^2=2(\epsilon-3)^2\Big(\epsilon-\sum_i\frac{(\alpha_i\beta_i)^2}{2}\Big)+2\sqrt2\sum_i\alpha_i^3\beta_i x_i y_i^2=0,\label{eq:multiCond4}}
we obtain the new condition that the combination $\sum_i\alpha_i^3\beta_i x_i y_i^2$ is a constant in time  
\dis{\frac{d}{Hdt}\sum_i\alpha_i^3\beta_i x_i y_i^2=&-\frac{3}{\sqrt2}(\epsilon-3)^3\Big(\epsilon-\sum_i\frac{(\alpha_i\beta_i)^2}{2}\Big)-(\epsilon-3)\sum_i\frac{\alpha_i^4\beta_i^2}{\sqrt2}y_i^2
\\
&-\sqrt2\sum_i\alpha_i^4\beta_i(y_i^4-2x_i^2y_i^2)
\\
=&0. \label{eq:multiCond5}}
Then conditions \eqref{eq:multiCond3}, \eqref{eq:multiCond4}, and \eqref{eq:multiCond5}, together with $\sum_i(x_i^2+y_i^2)=\epsilon$ and $\sum_i\frac{\alpha_i\beta_i}{\sqrt2}x_i=\epsilon$ uniquely determine the values of $(x_1, x_2, y_i, y_2, \epsilon)$. 
In particular, each of $x_i$ and $y_i$ is a constant in time so we do not need to take further time derivatives to obtain additional conditions.
For example, in Type IIB string compactification, the values of $(\alpha_i, \beta_i)$ for the volume modulus and the axio-dilaton are given by $(\sqrt{\frac23}, 3)$ and $(\sqrt2, 1)$, respectively, from which five conditions read
\dis{&(1)\quad x_1^2+y_1^2+x_2^2+y_2^2=\epsilon,
\\
&(2)\quad \sqrt3 x_1+ x_2=\epsilon,
\\
&(3)\quad (\epsilon-3)(\epsilon-4)=2(y_1^2+y_2^2),
\\
&(4)\quad -\frac14(\epsilon-3)^2(\epsilon-4)=\frac{1}{\sqrt3}x_1y_1^2+x_2y_2^2,
\\
&(5)\quad -\frac{1}{\sqrt2}(\epsilon-3)^2(\epsilon-4)-\frac{4}{\sqrt2}(\epsilon-3)(y_1^2+y_2^2)+4\sqrt2(\epsilon-3)\Big(\frac{1}{\sqrt3}x_1y_1^2+x_2y_2^2\Big)
\\
&\quad\quad\quad -4\sqrt2\Big(\frac13(y_1^4-2x_1^2y_1^2)+(y_2^4-2x_2^2y_2^2)=0.}
Solving these conditions, we obtain following values of $(x_1, x_2, y_i, y_2, \epsilon)$ : 
\dis{&({\rm A})\quad \epsilon=2\quad{\rm at}\quad x_1=\frac{\sqrt3}{2},\quad y_1=\frac{\sqrt3}{2}\quad x_1=\frac12,\quad y_2=\frac12,
\\
&({\rm B})\quad\epsilon=\frac{11}{5}\quad{\rm at}\quad x_1=\frac{2\sqrt3}{5},\quad y_1=\frac{3\sqrt2}{5}\quad x_1=1,\quad y_2=0,}
which correspond to Case 1, and
\dis{&({\rm C})\quad\epsilon=3\quad{\rm at}\quad x_1=\frac{\sqrt3}{2},\quad y_1=0\quad x_1=\frac32,\quad y_2=0,
\\
&({\rm D})\quad\epsilon=3\quad{\rm at}\quad x_1= \sqrt3,\quad y_1=0\quad x_1=0,\quad y_2=0,\label{eq:boundstab}}
which correspond to Case 3.
To see the stability of the fixed points, we consider the variations $\frac{d x_i}{Hdt}$ and $\frac{d y_i}{Hdt}$ at linear order in $(\delta x_i, \delta y_i)$   given by
\dis{({\rm A})\quad  \quad &\frac{dx_1}{Hdt}=-\frac52\delta x_1-\frac{\sqrt3}{2}\delta x_2-\frac72\delta y_1-\frac{\sqrt3}{2}\delta y_2,
\\
&\frac{dy_1}{Hdt}= \frac52\delta x_1+\frac{\sqrt3}{2}\delta x_2+\frac32\delta y_1+\frac{\sqrt3}{2}\delta y_2,
\\
&\frac{dx_2}{Hdt}=-\frac{\sqrt3}{2}\delta x_1-\frac32\delta x_2-\frac{\sqrt3}{2}\delta y_1-\frac52\delta y_2,
\\
&\frac{dy_2}{Hdt}= \frac{\sqrt3}{2}\delta x_1+\frac32\delta x_2+\frac{\sqrt3}{2}\delta y_1+\frac12\delta y_2, }
\dis{({\rm B})\quad  \quad &\frac{dx_1}{Hdt}=-\frac{56}{25}\delta x_1-\frac{6\sqrt3}{5}\delta x_2-\frac{38\sqrt6}{25}\delta y_1,
\\
&\frac{dy_1}{Hdt}= \frac{22\sqrt6}{25}\delta x_1+\frac{6\sqrt2}{5}\delta x_2+\frac{36}{25}\delta y_1,
\\
&\frac{dx_2}{Hdt}=-\frac45\delta x_2,
\quad
\frac{dy_2}{Hdt}= \frac65\delta y_2, }
\dis{({\rm C})\quad  \quad &\frac{dx_1}{Hdt}=-\frac32\delta x_1-\frac{3\sqrt3}{2}\delta x_2,
\quad
\frac{dy_1}{Hdt}=  \delta y_1,
\\
&\frac{dx_2}{Hdt}=\frac{\sqrt3}{2}\delta x_1+\frac32\delta x_2,
\quad
\frac{dy_2}{Hdt}= 3\delta y_2, }
\dis{({\rm D})\quad  \quad &\frac{dx_1}{Hdt}=0,
\quad
\frac{dy_1}{Hdt}=  2\delta y_1,
\\
&\frac{dx_2}{Hdt}=-2\sqrt3  \delta x_1,
\quad
\frac{dy_2}{Hdt}= 0. }
Then we find that for the cases (B), (C), and (D), at least one time variation of $y_i$ is given by the form $\frac{dy_i}{Hdt}=\gamma \delta y_i$ with some positive $\gamma_i$, which indicates that the fixed point is not stable with respect to the direction of $y_i$.
Moreover, for the case (A), which corresponds to Case 1 in Sec. \ref{Sec:Dynamics}   with $\epsilon_0=\frac{3\beta}{2+\beta}$, the fixed point is on the meridian of the ellipsoid $\epsilon=\sum_i\frac{\alpha_i\beta_i}{\sqrt2}x_i$, which is  written as $(x_1-\frac{\sqrt3}{2})^2+(x_2-\frac12)^2+y_1^2+y_2^2=1$.
In the language of Case 1 in Sec. \ref{Sec:Dynamics}, the fixed point value of $x$ is not larger than the value of $x$ at the center of the curve $x^2+y^2-\frac{\alpha\beta}{\sqrt2}x=0$.
This implies that the direction like the arrow (a) or (c) in Fig. \ref{Fig:abss6sta} (2) does not exist hence the fixed point is not stable.

Meanwhile, when $\epsilon=3$ but the combination $\sum_i\frac{\alpha_i\beta_i}{\sqrt2}x_i$ is not necessarily identified with $\epsilon$ (the generalization of Case 2), time variations of $x_i$ and $y_i$ satisfy
\dis{\frac{d x_i}{Hdt}=-\sqrt2 \alpha_i y_i^2,\quad \frac{d y_i}{Hdt}=\sqrt2 \alpha_i x_i y_i.}
From this, we immediately find that for each pair of $(u_i, v_i)$,  the combination $x_i^2+y_i^2$  is a constant in time.
In this case, even though $x_i$ and $y_i$ vary in time, as far as they are on the curve $\epsilon=3$ the value of $\epsilon$ does not change in time.

 \section{Conclusions}
\label{sec:conclusion}

 We studied the  fixed point values of the slow-roll parameter $\epsilon$ and their stability when the axion as well as the saxion  moves in the quintessence model.
 Even though the potential at tree level does not depend on the axion due to the shift symmetry, the axion couples to the saxion as well as the background geometry.
 As a result, the axion can move as  the saxion rolls down the runaway potential.
 Then  the axion kinetic energy contributes to the vacuum energy density and $\epsilon$ in  nontrivial ways at the fixed point where  the slow-roll approximation does not hold.
 Even in this case, just like the case in which only the saxion moves, the   value of $\epsilon$ is restricted to be smaller than $3$, which is imposed by the positivity of the potential, and its properties at the fixed point   depend on   $\alpha$ and $\beta$.
We also find that the fixed point is in general   not stable with respect to the time evolution from all directions in the field space. 
  To see this more concretely, we consider the volume modulus and the axio-dilaton, which are the essential ingredients of the string compactification and couple to the potential universally.
  When one of them is fixed, there exist some partially stable fixed points and   if the value of $\epsilon$ is given by $3$, the largest value imposed by the positivity of the potential, it is not changed even if the values of $x$ and $y$ vary in time.
  If both of them are allowed to move, we find that the stable fixed point does not exist.  
  Our analysis shows that the addition of the axion in the stringy quintessence model can exhibit different feature from the simplest model in which only the saxion is dynamical.


\appendix



\renewcommand{\theequation}{\Alph{section}.\arabic{equation}}


\subsection*{Acknowledgements} 

This work is motivated by Filippo Revello's question about the author's previous work \cite{Seo:2024fki}.


\end{document}